# Brain on the 3D Visual Art through Virtual Reality; Introducing Neuro-Art in a Case Investigation

Ali-Mohammad Kamali[1,2,3], Mohammad Taghi Najafi[4,5], Mohammad Nami[1,2,3,6*]

[1]Department of Neuroscience, School of Advanced Medical Sciences and Technologies, Shiraz University of Medical Sciences, Shiraz, Iran
[2]DANA Brain Health Institute, Iranian Neuroscience Society-Fars Branch, Shiraz, Iran
[3]Neuroscience Laboratory, NSL (Brain, Cognition and Behavior), Department of Neuroscience, School of Advanced Medical Sciences and Technologies, Shiraz University of Medical Sciences, Shiraz, Iran
[4]Payame Noor University, Dubai, United Arab Emirates
[5] Harmony Healing Center and Aspetto Research Center, Dubai, United Arab Emirates
[6]Academy of Health, Senses Cultural Foundation, Sacramento, CA, USA

## Abstract

The reciprocal impact of applied neuroscience and cognitive studies on humanities has been extensive and growing over the past 30 years of research. Studies on neuroaesthetics have provided novel insights in visual arts, music as well as abstract and dramatic art. Neuro-Art is an experimental concept in applied neuroscience where scientists can study the mechanistic pathways involved for instance in visual art through which creativity and artistic capacity might receive further empowerment. Based on the existing evidence, at least 3 large-scale brain networks are involved simultaneously when one is submitted to a creativity-related task. The question whether the key brain regions involved in visual art creativity can be identified and receive neuromodulation to get empowered prompted us to perform the present case investigation. Virtual reality and functional quantitative electroencephalography upon 2- vs 3-dimentional painting were employed to study cortical neurodynamics in a professional painting artist.

## 1. Background and hypothesis

Art is indeed regarded as a feature of all societies and valued by the people (Duncum, 2003). Since art is an example of human activity, it is believed to depend on and follow the laws of the brain (Zeki, 2002).
When we explore elements of our inner representations, we use visual qualities such as color, brightness, view angle and clarity as well as acoustic features including pitch, meaning or resemblance, melody and rhythm. This is the case in ordinary people whereas artists with outstanding creative capacity may possess heightened awareness of specific qualia and employ their surrounding objects in a way that they stimulate these states by intuition or by trial and error(Andujar, Crawford, Nijholt, Jackson, & Gilbert, 2015).

For instance, when someone hears the word 'cake', he or she might visualize the image of the cake, imagine its taste, perhaps smell one or feel its texture. Some other individuals may however visualize the written word 'c-a-k-e' and think about possible recipes. The inter-individual differences would in some ways be attributed to our memories and internal future representations. On the other hand, anyone may have a dominant lead representational system through the sense which his or her mind first uses to trawl a representation. The more visual, auditory and kinesthetic data we process as we encounter a motif, the stronger and more compelling will be the qualia we perceive. The predominance of one or more senses in capturing representations and the related neurobehavioral correlates have been the basis of neurolinguistics programing (Ahmad, 2017; Damiano, Kercel, Tucker, & Brown-VanHoozer, 2000).

Art is a fine skill acquired through observation, learning and continued practice, meanwhile is typically generated by the human creativity (West, 2000; Zimmerman, 2009). Today's advances in neuroscience and related technologies have enabled us to investigate brain processes in relation to artistic performance. Evidence form the existing studies have postulated that highly creative individuals do not acquire hemispheric dominance(Hyman, 2010). While the right hemisphere is more involved in divergent thinking and performs as the hub for imagination and contemplation, a creative artist would involve the left hemisphere to maintain the balance (Zaidel, 2013). According to earlier investigations, a range of functional brain networks are known to be involved in the practice of visual art (Cela-Conde et al., 2013). Although there have been much discussions on the nature of art and how it is formulated in our brain there has been no satisfactory conclusion so far. It appears that neural scientists are still at the prime of the journey to unveil the neural basis of the visual art (Cela-Conde et al., 2013).

The influence of neuroscience and cognitive studies on humanities and vice versa have been substantial and growing over the past three decades. In other words, there has been a notable discourse between the concept of neuroaesthetics and neurocognitive paradigms with regards to the inquiry of art and humanities. As such, neuroaesthetic studies and related applications have led to new perspectives in visual arts, music as well as abstract and dramatic art (Jameson, 2012; Onians, 2018).

With the view of the fact that the brain holds the capacity of forming new connections, it is known to modify itself through formation of new pathways, hence regarded to be plastic. The concept of neural plasticity informs that the functional capacity and neurodynamics of the brain can receive favorable changes through practice and the use of neuromodulation techniques (Alcami & Pereda, 2019; Baker, Rao, & Mizumori, 2019; Morris et al., 2019; Soekadar, Herring, & McGonigle, 2016). Since art is shown to stimulate the brain and induce aesthetic or emotional perceptions with consequent plastic changes, it appears worthwhile to encourage more research on the concept of neuro-art, where art and neuroscience intermingle (Alcami & Pereda, 2019; Baker et al., 2019; Nami & Ashayeri, 2011).

In an earlier research, we described where neuroscience and art might interweave. We came across a contemporary painting artist who turned to be a synesthetic. He claimed to be hearing the sound of colors. In other words, when he was listening to a favorite music he could turn the notes, pitches and melodies into shades of given colors and robust moves on the canvas. When examined in a neuroscience laboratory using the quantitative electroencephalogram (qEEG), some event-related desynchronizations were documented in his acoustic cortices while he was viewing his synesthetic experience artworks ( for example a Schubert's classical music turned into a painting) while this was not the case upon other stimuli presentations (Nami & Ashayeri, 2011).

The above typical example together with similar scenarios and open questions have prompted research groups to tap into the neural dynamics of creative visual art (Alain et al., 2019; Luft, Zioga, Banissy, & Bhattacharya, 2019). The question of how the brain puts together creativity in art has been intriguing and became a substrate of several research projects on the interface of art and neuroscience (Neuro-Art) (Alain et al., 2019; Cheung, Law, Yip, & Wong, 2019; Ho Tiu et al., 2019; Luft et al., 2019; Mukunda et al., 2019; Siler, 2015). This has been dealt with through cognitive, behavioral, neurological, psychological and humanities/social science standpoints. The concept of Neuro-Art has been involved in systematic investigations to explain neural basis of visual and auditory perception, emotional reaction to art and the cognitive content involved as well as the basis of creativity and aesthetic experience (Nami & Ashayeri, 2011).

In a recent study, Roger E. Beaty et al. investigated the individual creative ability in relation with brain functional connectivity using functional MRI (fMRI) scans (Beaty et al., 2014). Their findings suggested that among several large-scale functional brain networks, the whole-brain network associated with high-creative ability includes cortical pivots within three networks namely default, salience, and executive networks. The findings potentially highlights the notion that highly creative people including artists (here, visual arts) retain an increased capacity to simultaneously involve such three large-scale brain networks when creating artworks (Beaty et al., 2014).

The activity of the large-scale brain networks can be practically measured using the fMRI technique (Brueggen et al., 2017; Zhang et al., 2013). The default mode network (DMN) is a network of interrelating brain areas which show heightened activity when an individual is not focused on the surrounding world. This network's activity predominates when the brain is considered to be in resting state (Vessel, Starr, & Rubin, 2012, 2013). On the other hand, the salience network comprises brain areas to select which stimuli we selectively attend to. The insular cortex plays a central role in the salience network to target behaviorally significant stimuli and to direct the brain's neural mechanisms in response to such stimuli(Hilland, Landro, Harmer, Maglanoc, & Jonassen, 2018; Janes et al., 2018). In addition, the main components of the central executive network are known to be the prefrontal cortex and posterior parietal cortex (Beaty et al., 2014) (Figure 1, Panel A).

Based on the current source density (CSD) analysis of the qEEG signals, there are some key cortical areas known correspond to the hotspots involved in the interface of the three above mentioned large-scale brain networks (DMN, SN and CEN)(Beaty et al., 2014). According to the international 10-20 EEG system, these cortical areas include but not restricted to FPz, Pz, F3, F4, T7, T8, P5 and P6 (Figure 1, Panel B).

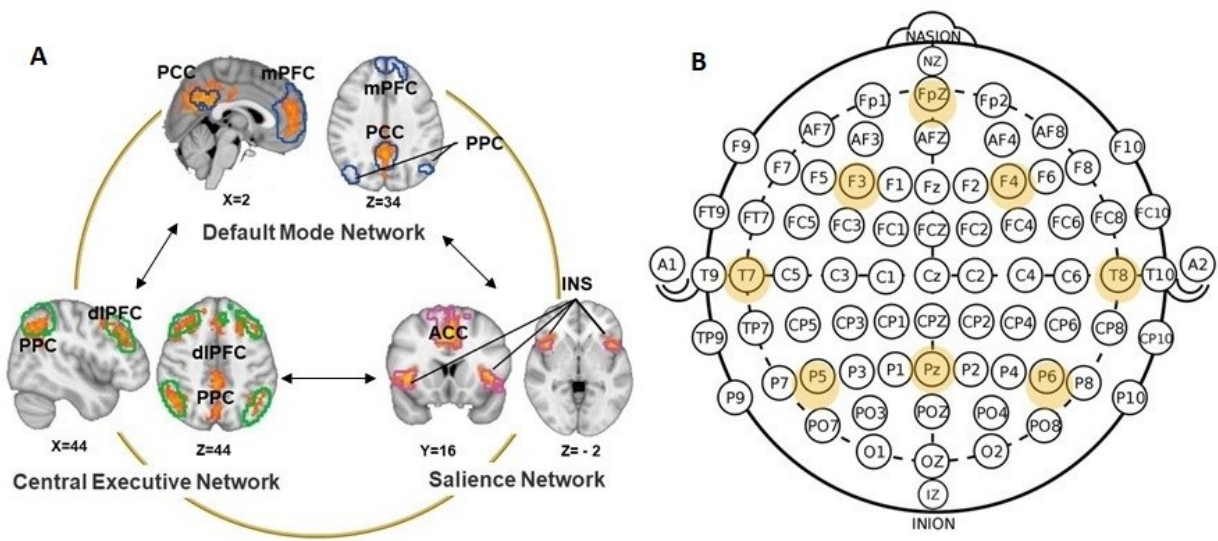

Figure 1. The salience, executive control, and default mode networks. Colored outlines depict the a priori regions of interest that defined each network (A). The international 10-12 EEG system whereby channels corresponding to the involved areas with the interface of the 3 networks when simultaneously activated are marked (B). Figure adapted from Beaty et al. 2014: 10.1016/j.neuropsychologia.2014.09.019 under CC license.

Painting artist typically create their art works in a two-dimensional (2D) space i.e. paper, canvas etc. Meanwhile, the advent of a three-dimensional (3D) drawing environment in virtual reality by Google ("Tilt Brush by Google,") has prompted us to investigate whether the 3D space drawing while the artist is immersed in virtual reality (VR) could potentiate the activity of DMN, SN and CEN even more than the 2D. If that held true, one could have of using behavioral approaches using VR plus neuromodulation techniques (e.g. transcranial electrical stimulation-tES) upon the involved cortical hotspots to trigger, improve or empower creativity in artists. This is based on the fundamental theories and applied neurotechnological approaches with the growing evidence supporting neuroplasticity in the brain.
The above hypothesis stirred us to pursue a case study on a painting artist.

## 2. Case study

Mr. A.N. was a 50 year-old painting artist with over 25 years of professional experience and more than 10 local/international solo and group art-exhibitions in his career. He was healthy, taking no medications, non-smoker, non-alcoholic with insignificant past medical, surgical and neuropsychiatric history. After he gave a written informed consent to enter our study, he was submitted to the qEEG examination using the Mitsar monopolar 21-channel EEG device and electrocap over three consecutive days. All examinations were done in our electrophysiology unit at Dana Brain health institute.

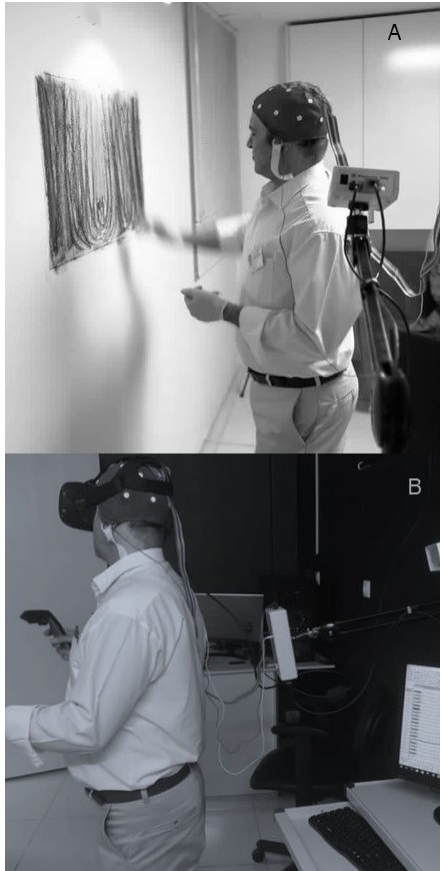

**Figure 2.** The 2D- and 3D-painting task concurrent with qEEG (A and B). The 3D-painting task involved the VR setup where the artist wore the VR goggles using both hands to hold gadgets. The right hand gadget functioned as the brush and the left hand gadget was the color and effects toolbox.

On each examination day, it was confirmed that his last night's sleep quality was equal to or above 7 (from the scale of 0-10, 10 being the most refreshing sleep in a visual analogue scale-VAS), his stress and anxiety scale score was below 15 (based on the Beck's Anxiety Inventory-BAI) and he has had no psychological unpleasant incident, and no caffeine in the morning. The experiments were done at 10:00 am -12:00 PM each day. All measures were possibly taken to unify the controlled laboratory conditions upon three consecutive days when he was examined by qEEG in resting state, 30 min [Day-1]; qEEG in resting state (30 min) followed by qEEG upon 2D-painting using a paper sheet and crayons, 30 min [Day-2]; and qEEG in resting state (30 min) followed by qEEG upon 3D-painting using the virtual reality setup (HTC Vive, Dual AMOLED 3.6" diagonal screen with 90 Hz refresh rate) and the Google's tilt Brush software, 30 min [Day-3].

| Default-Mode Network | | | Salience Network | | | Central Executive Network | | |
|---|---|---|---|---|---|---|---|---|
| Resting state | 2D-painting | 3D-painting | Resting state | 2D-painting | 3D-painting | Resting state | 2D-painting | 3D-painting |
| Key cortical areas involved based on the international 10-20 system <br> FPz, Pz, F3, F4, T7, T8, P5, P7 | | | | | | | | |
| Brodmann areas with maximal CSD loading at SMR band witch (13 Hz) <br> 10, 11, 30, 35 ( Rt and Lt) | | | Brodmann areas with maximal CSD loading at SMR band witch (13 Hz) <br> 10, 13, 25,22, 30 ( Rt and Lt) | | | Brodmann areas with maximal CSD loading at SMR band witch (13 Hz) <br> 10, 47, 11 ( Rt and Lt) | | |

**Figure 3.** Summary of the study protocol. CSD: Current Source Density, SMR: Sensory-Motor Rhythm.

For the VR experience, he wore the EEG cap and put on the VR goggle using both hands to hold gadgets where the right hand gadget functioned as the brush and the left hand gadget was the color and effects toolbox (Figure2).
The study design is illustrated in Figure 3.

When data acquisitions were concluded, EEG signals were preprocessed using the EEG-Lab in MATLAB to denoise signals from motion and eye-movement artifacts. Later, the EEG time series were analyzed using the Neuroguide software and Neuronavigator (Applied Neuroscience, USA, v.3.0.5, 2018) to extract data including absolute power, relative power, amplitude asymmetry and connectivity/coherence z-scores of the frequency band spectra for comparative analyses across states (Matusevich, Ruiz, & Vairo, 2002).

Based on the above, there were three states (resting state, 2D-painting and 3D-painting) in which qEEG signals were comparatively analyzed with regard to the CSD gain in key cortical areas potentially representing DMN, SN and CEN (FPz, Pz, F3, F4, T7, T8, P5 and P6) ( Figure 1).
The built-in Neurostat function in Neuroguide was employed for the inter-state data comparisons and statistical analyses (Ogura et al., 2009).

### *3. Results*

The comparative analyses of findings on the CSD gain in DMN qEEG upon resting state versus 2D- and 3D-painting are outlined in Table 1. It turned out that the peak center value for DMN foci declined in 2D-painting as compared to the resting state. Meanwhile, such values were notably higher in 3D-painting through VR inasmuch as center values for DMN's absolute power at 13 Hz spectral band was higher not only than the same in 2D-painting but also the resting state.

This indicates that the CSD at 13 Hz [within the range of sensory-motor rhythm (SMR) frequency] in the key cortical areas corresponding to the DMN was higher during the 3D-painting task as compared to the other two states. Taking into consideration the role of DMN in creativity and the fact that DMN shows a heightened activity when an individual is not focused on the outer world, the above finding suggests that 3D-painting in VR potentially stimulated the artist's capacity for his creative function at least at cortical level.

As demonstrated in Figures 4, the CSD peak values which are shown in color spectra were found to correspond to hot colors in key areas known to represent the interface of DMN, SN and CEN more in the 3D-painting task as compared to the other 2 states (left panel). In addition, as shown in Figure 2 (right panel), the coherence z-scores at 13 Hz amongst the Brodmann areas in common within the interface of DMN, SN and CE (BAs 10-Right, 10-Left, 11-Right, 11-Left, 30 –Right, 30-Left, 35-Right and 35-Left) gained higher values upon 3D-painting task as compared to the other 2 states.

**Table 1.** Mean center values for current source density at areas predominantly involved in DMN dynamics. The comparative changes among states (i.e. resting state, 2D-painting and 3D-painting tasks) are summarized. Data suggest a notable change in CSD in respective areas within the DMN mainly in the 3D- vs 2D- painting task.

| | Brodmann Area | Coordinate and anatomical name | Mean Value Min Value Max Value | Δ Mean value 2D painting vs. Rs, 13 Hz | Δ Mean value 3D painting vs. Rs, 13 Hz | Δ Mean value 3D vs. 2D painting, 13 Hz |
|---|---|---|---|---|---|---|
| **RS DMN** | 10 Right | X = 14, Y = -57, Z = -9 Frontal_Sup_Medial_R | 5.08333 -0.850301 7.56911 | N/A | N/A | N/A |
| | 10 Left | X = -13, Y = 58, Z = 9 Frontal_Sup_Medial_L | 5.15698 -0.850301 7.56911 | | | |
| | 30 Right | X = 14, Y = 37, Z = 13 Fusiform_R | 4.72588 -0.850301 7.56911 | | | |
| | 30 Left | X = -15, Y = -37, Z = -13 Fusiform_L | 4.345433 -0.850301 7.56911 | | | |
| | 35 Right | X = 14, Y = -13, Z = -24 Parahippocampal_R | 5.24507 -0.850301 7.56911 | | | |
| | 35 Left | X = -16, Y = -13, Z = -25 Parahippocampal_L | 5.97822 -0.850301 7.56911 | | | |
| **2D Painting-DMN** | 10 Right | See above | 4.12303 -0.654565 6.23321 | -0.9603 | N/A | N/A |
| | 10 Left | | 3.83345 -0.654565 6.23321 | -1.32248 | | |
| | 30 Right | | 3.23454 -0.654565 6.23321 | -1.49134 | | |
| | 30 Left | | 3.02221 -0.654565 6.23321 | -1.323223 | | |
| | 35 Right | | 3.98788 -0.654565 6.23321 | -1.25719 | | |
| | 35 Left | | 4.03454 -0.654565 6.23321 | -1.25719 | | |
| **3D Painting-DMN** | 10 Right | See above | 7.43345 -0.134545 9.76564 | N/A | +2.35012 | +3.31042 |
| | 10 Left | | 7.87676 -0.134545 9.76564 | | +2.71978 | +4.04226 |
| | 30 Right | | 6.56543 -0.134545 9.76564 | | +1.83955 | +3.33089 |
| | 30 Left | | 6.45658 -0.134545 9.76564 | | +2.111147 | +3.43437 |
| | 35 Right | | 8.23434 -0.134545 9.76564 | | +2.98927 | +4.24646 |
| | 35 Left | | 8.56765 -0.134545 9.76564 | | +2.58943 | +4.53311 |

Figure 5 illustrates the qEEG heat-map upon 3D-painting task showing that areas attributed to DMN ( FPz, FCz and Pz corresponding to the ventromedial prefrontal cortex and posterior parietal cortex, respectively) had a notable absolute power gain at theta and SMR band frequencies. There was likewise a predominance for amplitude asymmetry favoring frontocentral brain regions at theta and alpha band spectra. Meanwhile, no hemispheric dominance in terms of amplitude gain was observed across spectra.

## 4. Discussion

The growing insights into brain science and the advent of technologies including the functional magnetic, electrical and optical neuroimaging as well as neural stimulation techniques have assisted scholars to study brain processes upon artistic performance (Alain et al., 2019; Bolwerk, Mack-Andrick, Lang, Dorfler, & Maihofner, 2014; Liu & Miller, 2008; Pepperell, 2011).

Based on our findings and that of similar studies(Andujar et al., 2015; Beaty, Benedek, Kaufman, & Silvia, 2015; Beaty, Benedek, Silvia, & Schacter, 2016; Beaty et al., 2014), it appears that cortical regions which mainly correspond to the brain areas commonly involved within the interface of DMN, SN and CEN are positively loaded during the process of creativity in art. Our results highlighted that the activity within the DMN in 3D-painting using the *tilt Brush* function in VR may outweigh the DMN activity in 2D painting task and even resting state in a single case study.

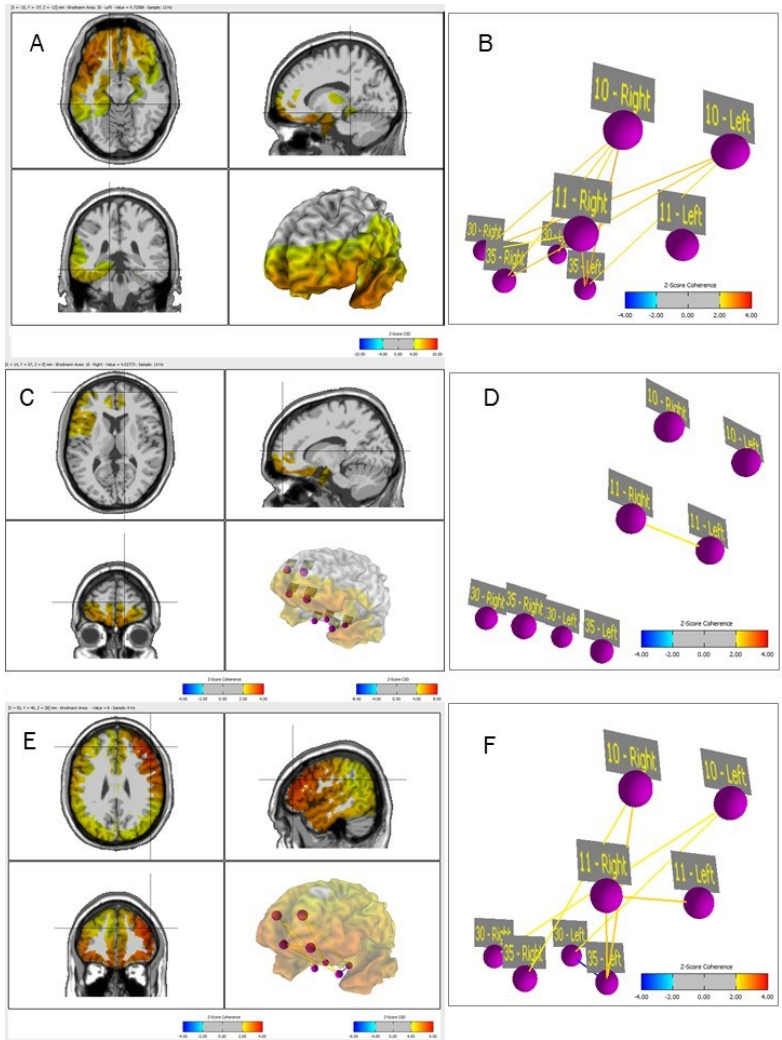

**Figure 4.** The CSD peak values shown in color spectra in the interface of DMN, SN and CEN in the 3D-painting task (left panel). The connectivity coherence z-scores at 13 Hz amongst the Brodmann areas in common within the interface of DMN, SN and CE upon 3D- painting task. CSD: Current Source Density, DMN: Default Mode Network, SN: Salience Network, CEN: Central Executive Network.

When this notion is further examined and approved in large-scale studies, one may consider pursuing cortical stimulation on such related cortical hotspots in artists aiming towards enhanced creativity capacity (Green et al., 2017; Ivancovsky, Kurman, Morio, & Shamay-Tsoory, 2018; Lucchiari, Sala, & Vanutelli, 2018; Mayseless & Shamay-Tsoory, 2015).

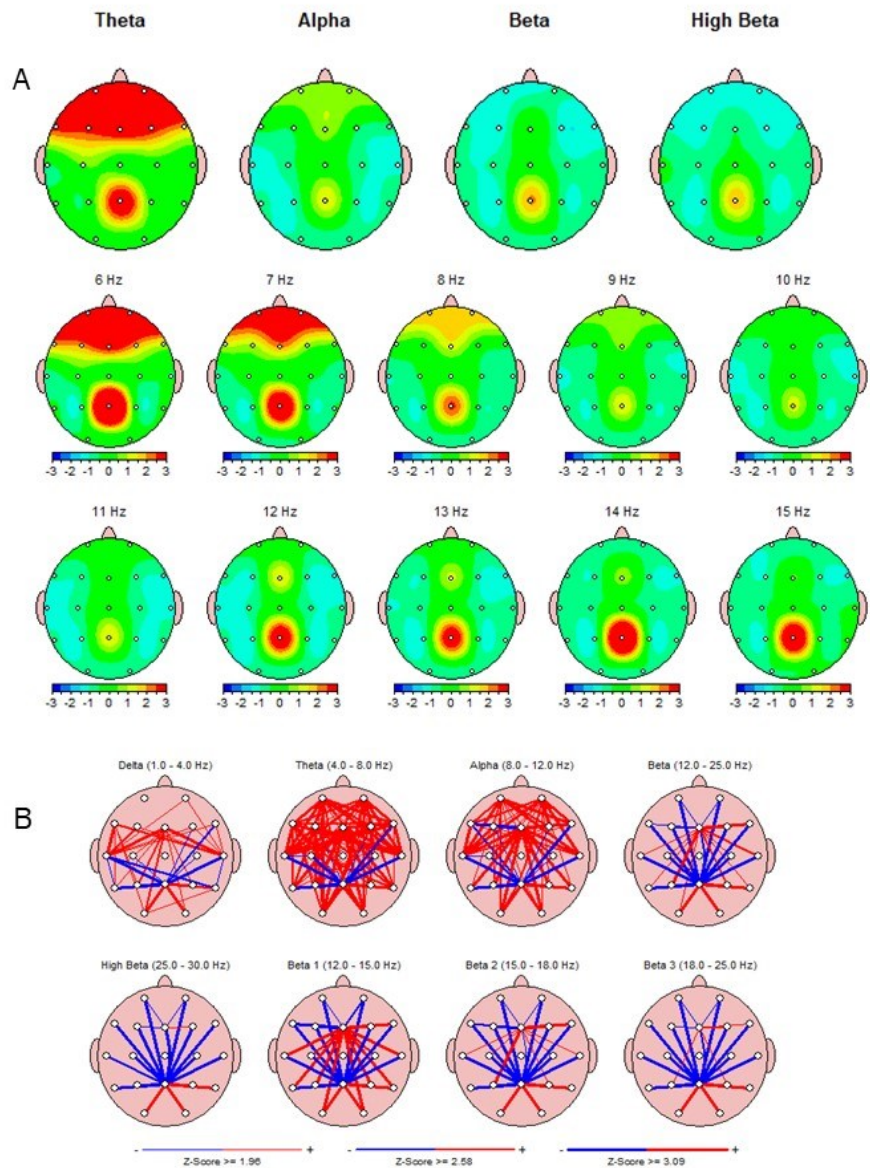

Figure 5. The qEEG heat-map upon 3D-painting task demonstrating a notable z-score absolute power gain at theta and SMR band frequencies. No hemispheric dominance in terms of amplitude gain was observed across spectra. SMR: Sensory-Motor Rhythm.

A range of novel neurotechnology tools with proper safety profiles including transcranial electrical stimulation (tES), repetitive transcranial magnetic stimulation (rTMS) and neurofeedback may be applied for the above purpose(Luft et al., 2019; Tatti, Rossi, Innocenti, Rossi, & Santarnecchi, 2016; Vicario & Nitsche, 2013). This is what popularly regarded as neurocognitive empowerment in the today context of applied neuroscience. The use of VR may provide an add-on value in terms of behavioral training to potentiate creativity among artists and trainees of visual art (Gerry, 2017).

The creativity arising from artist's brain is hypothesized to be linked to an intricate neural circuitry in which large-scale brain networks namely DMN, SN and CEN play a central part (Beaty et al., 2016; Beaty et al., 2014). In the context of creativity in visual art, each network prompts an activity in a particular part of the visual brain, the superior orbitofrontal, the mid-cingulate, and the anterior cingulate cortex suggesting functional organization where creative thinking and visual art performance overlap (Beaty et al., 2016; Beaty et al., 2014). In the sense of art experience, according to Barbur et al., all human experience is mediated through the brain and is not solely the product of the outside world. The more important the experience, the more it can reveal about the fundamental properties of the brain" (Barbur, Watson, Frackowiak, & Zeki, 1993).

Taken the inter-individual differences in creative capacity, cognitive style, thinking process and the extent of training in artistic experience, each brain may need to be well mapped before neuromodulation and behavioral training are applied towards the expected empowerment where art and creativity interweave.

## 5. Conclusion

The outcome of this case study suggests that when the artist's brain was involved in a visual art task to provoke further creativity (i.e. the 3D-painting task), the activity of DMN at the SMR frequency band gained a greater CSD than the 2D-painting task and resting state.

### *Authors' contribution*

The work is a product of the whole team; all members have contributed to the conception of the work.

# References

Ahmad, K. Z. (2017). *Improving Emotional Intelligence (EI) using Neuro Linguistic Programming (NLP) techniques.* Paper presented at the International Conference on Advances in Business, Management and Law (ICABML).

Alain, C., Moussard, A., Singer, J., Lee, Y., Bidelman, G. M., & Moreno, S. (2019). Music and Visual Art Training Modulate Brain Activity in Older Adults. *Front Neurosci, 13*, 182. doi:10.3389/fnins.2019.00182

Alcami, P., & Pereda, A. E. (2019). Beyond plasticity: the dynamic impact of electrical synapses on neural circuits. *Nat Rev Neurosci*. doi:10.1038/s41583-019-0133-5

Andujar, M., Crawford, C. S., Nijholt, A., Jackson, F., & Gilbert, J. E. (2015). Artistic brain-computer interfaces: the expression and stimulation of the user's affective state. *Brain-computer interfaces, 2*(2-3), 60-69.

Baker, P. M., Rao, Y., & Mizumori, S. J. Y. (2019). Transient Input-Specific Neural Plasticity in the Lateral Habenula Facilitates Learning. *Neuron, 102*(1), 1-3. doi:10.1016/j.neuron.2019.03.019

Barbur, J. L., Watson, J. D., Frackowiak, R. S., & Zeki, S. (1993). Conscious visual perception without V1. *Brain, 116 ( Pt 6)*, 1293-1302.

Beaty, R. E., Benedek, M., Kaufman, S. B., & Silvia, P. J. (2015). Default and Executive Network Coupling Supports Creative Idea Production. *Sci Rep, 5*, 10964. doi:10.1038/srep10964

Beaty, R. E., Benedek, M., Silvia, P. J., & Schacter, D. L. (2016). Creative Cognition and Brain Network Dynamics. *Trends Cogn Sci, 20*(2), 87-95. doi:10.1016/j.tics.2015.10.004

Beaty, R. E., Benedek, M., Wilkins, R. W., Jauk, E., Fink, A., Silvia, P. J., . . . Neubauer, A. C. (2014). Creativity and the default network: A functional connectivity analysis of the creative brain at rest. *Neuropsychologia, 64*, 92-98. doi:10.1016/j.neuropsychologia.2014.09.019

Bolwerk, A., Mack-Andrick, J., Lang, F. R., Dorfler, A., & Maihofner, C. (2014). How art changes your brain: differential effects of visual art production and cognitive art evaluation on functional brain connectivity. *PLoS One, 9*(7), e101035. doi:10.1371/journal.pone.0101035

Brueggen, K., Fiala, C., Berger, C., Ochmann, S., Babiloni, C., & Teipel, S. J. (2017). Early Changes in Alpha Band Power and DMN BOLD Activity in Alzheimer's Disease: A Simultaneous Resting State EEG-fMRI Study. *Front Aging Neurosci, 9*, 319. doi:10.3389/fnagi.2017.00319

Cela-Conde, C. J., García-Prieto, J., Ramasco, J. J., Mirasso, C. R., Bajo, R., Munar, E., . . . Maestú, F. (2013). Dynamics of brain networks in the aesthetic appreciation. *Proceedings of the National Academy of Sciences, 110*(Supplement 2), 10454-10461.

Cheung, M. C., Law, D., Yip, J., & Wong, C. W. Y. (2019). Emotional Responses to Visual Art and Commercial Stimuli: Implications for Creativity and Aesthetics. *Front Psychol, 10*, 14. doi:10.3389/fpsyg.2019.00014

Damiano, B., Kercel, S. W., Tucker, R., & Brown-VanHoozer, S. A. (2000). *Recognizing a Voice from its Model.* Paper presented at the Smc 2000 conference proceedings. 2000 ieee international conference on systems, man and cybernetics.'cybernetics evolving to systems, humans, organizations, and their complex interactions'(cat. no. 0.

Duncum, P. (2003). The theories and practices of visual culture in art education. *Arts Education Policy Review, 105*(2), 19-25.

Gerry, L. J. (2017). Paint with Me: Stimulating Creativity and Empathy While Painting with a Painter in Virtual Reality. *IEEE Trans Vis Comput Graph, 23*(4), 1418-1426. doi:10.1109/TVCG.2017.2657239

Green, A. E., Spiegel, K. A., Giangrande, E. J., Weinberger, A. B., Gallagher, N. M., & Turkeltaub, P. E. (2017). Thinking Cap Plus Thinking Zap: tDCS of Frontopolar Cortex Improves Creative Analogical Reasoning and Facilitates Conscious Augmentation of State Creativity in Verb Generation. *Cereb Cortex, 27*(4), 2628-2639. doi:10.1093/cercor/bhw080

Hilland, E., Landro, N. I., Harmer, C. J., Maglanoc, L. A., & Jonassen, R. (2018). Within-Network Connectivity in the Salience Network After Attention Bias Modification Training in Residual Depression: Report From a Preregistered Clinical Trial. *Front Hum Neurosci, 12*, 508. doi:10.3389/fnhum.2018.00508

Ho Tiu, C., Asfour, L., Jakab, M., Tomlin, H., Griffiths, C. E. M., & Young, H. S. (2019). An art-based visual literacy training course to enhance clinical skills in dermatology trainees. *J Eur Acad Dermatol Venereol*. doi:10.1111/jdv.15588

Hyman, J. (2010). *Art and neuroscience.* Paper presented at the Beyond mimesis and convention.

Ivancovsky, T., Kurman, J., Morio, H., & Shamay-Tsoory, S. (2018). Transcranial direct current stimulation (tDCS) targeting the left inferior frontal gyrus: Effects on creativity across cultures. *Soc Neurosci*, 1-9. doi:10.1080/17470919.2018.1464505

Jameson, E. (2012). merging art and neuroscience. *Ann Neurosci, 19*(2), 60. doi:10.5214/ans.0972.7531.12190202

Janes, A. C., Gilman, J. M., Frederick, B. B., Radoman, M., Pachas, G., Fava, M., & Evins, A. E. (2018). Salience network coupling is linked to both tobacco smoking and symptoms of attention deficit hyperactivity disorder (ADHD). *Drug Alcohol Depend, 182*, 93-97. doi:10.1016/j.drugalcdep.2017.11.005

Liu, A., & Miller, B. L. (2008). Visual art and the brain. *Handb Clin Neurol, 88*, 471-488. doi:10.1016/S0072-9752(07)88024-9

Lucchiari, C., Sala, P. M., & Vanutelli, M. E. (2018). Promoting Creativity Through Transcranial Direct Current Stimulation (tDCS). A Critical Review. *Front Behav Neurosci, 12*, 167. doi:10.3389/fnbeh.2018.00167

Luft, C. D. B., Zioga, I., Banissy, M. J., & Bhattacharya, J. (2019). Spontaneous Visual Imagery During Meditation for Creating Visual Art: An EEG and Brain Stimulation Case Study. *Front Psychol, 10*, 210. doi:10.3389/fpsyg.2019.00210

Matusevich, D., Ruiz, M., & Vairo, M. C. (2002). [QEEG and brain mapping. Historial develoment, clinical practices and epistemological issues]. *Vertex, 13*(49), 198-204.

Mayseless, N., & Shamay-Tsoory, S. G. (2015). Enhancing verbal creativity: modulating creativity by altering the balance between right and left inferior frontal gyrus with tDCS. *Neuroscience, 291*, 167-176. doi:10.1016/j.neuroscience.2015.01.061

Morris, J. H., Kelly, C., Joice, S., Kroll, T., Mead, G., Donnan, P., . . . Williams, B. (2019). Art participation for psychosocial wellbeing during stroke rehabilitation: a feasibility randomised controlled trial. *Disabil Rehabil, 41*(1), 9-18. doi:10.1080/09638288.2017.1370499

Mukunda, N., Moghbeli, N., Rizzo, A., Niepold, S., Bassett, B., & DeLisser, H. M. (2019). Visual art instruction in medical education: a narrative review. *Med Educ Online, 24*(1), 1558657. doi:10.1080/10872981.2018.1558657

Nami, M., & Ashayeri, H. (2011). Where Neuroscience and Art Embrace; the Neuroaesthetics. *Basic and Clinical Neuroscience, 15*(2), 6-11.

Ogura, T., Hida, K., Masuzuka, T., Saito, H., Minoshima, S., & Nishikawa, K. (2009). An automated ROI setting method using NEUROSTAT on cerebral blood flow SPECT images. *Ann Nucl Med, 23*(1), 33-41. doi:10.1007/s12149-008-0203-7

Onians, J. (2018). Art, the visual imagination and neuroscience: The Chauvet Cave, Mona Lisa's smile and Michelangelo's terribilita. *Cortex, 105*, 182-188. doi:10.1016/j.cortex.2017.10.009

Pepperell, R. (2011). Connecting Art and the Brain: An Artist's Perspective on Visual Indeterminacy. *Front Hum Neurosci, 5*, 84. doi:10.3389/fnhum.2011.00084

Siler, T. (2015). Neuroart: picturing the neuroscience of intentional actions in art and science. *Front Hum Neurosci, 9*, 410. doi:10.3389/fnhum.2015.00410

Soekadar, S. R., Herring, J. D., & McGonigle, D. (2016). Transcranial electric stimulation (tES) and NeuroImaging: the state-of-the-art, new insights and prospects in basic and clinical neuroscience. *Neuroimage, 140*, 1-3. doi:10.1016/j.neuroimage.2016.08.020

Tatti, E., Rossi, S., Innocenti, I., Rossi, A., & Santarnecchi, E. (2016). Non-invasive brain stimulation of the aging brain: State of the art and future perspectives. *Ageing Res Rev, 29*, 66-89. doi:10.1016/j.arr.2016.05.006

Tilt Brush by Google.   Retrieved from https://www.tiltbrush.com/

Vessel, E. A., Starr, G. G., & Rubin, N. (2012). The brain on art: intense aesthetic experience activates the default mode network. *Front Hum Neurosci, 6*, 66. doi:10.3389/fnhum.2012.00066

Vessel, E. A., Starr, G. G., & Rubin, N. (2013). Art reaches within: aesthetic experience, the self and the default mode network. *Front Neurosci, 7*, 258. doi:10.3389/fnins.2013.00258

Vicario, C. M., & Nitsche, M. A. (2013). Non-invasive brain stimulation for the treatment of brain diseases in childhood and adolescence: state of the art, current limits and future challenges. *Front Syst Neurosci, 7*, 94. doi:10.3389/fnsys.2013.00094

West, M. (2000). State of the art: creativity and innovation at work. *Psychologist, 13*(9), 460-464.

Zaidel, D. W. (2013). Split-brain, the right hemisphere, and art: Fact and fiction *Progress in brain research* (Vol. 204, pp. 3-17): Elsevier.

Zeki, S. (2002). Inner vision: An exploration of art and the brain.

Zhang, G., Zhang, H., Li, X., Zhao, X., Yao, L., & Long, Z. (2013). Functional alteration of the DMN by learned regulation of the PCC using real-time fMRI. *IEEE Trans Neural Syst Rehabil Eng, 21*(4), 595-606. doi:10.1109/TNSRE.2012.2221480

Zimmerman, E. (2009). Reconceptualizing the role of creativity in art education theory and practice. *Studies in Art Education, 50*(4), 382-399.